\documentclass[12pt]{iopart}
\usepackage{amssymb}
\usepackage{amsbsy}
\usepackage{graphicx}
\usepackage{epstopdf}
\usepackage{color}
\usepackage{graphicx,sidecap}
\usepackage{bm}

\usepackage[colorlinks,bookmarks=false,citecolor=blue,linkcolor=red,urlcolor=blue]{hyperref}

\catcode`@=11 
\renewcommand\tableofcontents{%
  \section*{\contentsname}%
  \@starttoc{toc}%
}
\catcode`@=12

\newcommand{\bra}[1]{\langle\left.{#1}\right|}
\newcommand{\ket}[1]{\left|{#1}\right.\rangle}

\def\be{\begin{equation}}
\def\ee{\end{equation}}
\def\u{\uparrow}

\begin{document}

\setlength{\parindent}{0pt}

\title{Entanglement spectrum of the Heisenberg XXZ chain near
the ferromagnetic point}

\author{Vincenzo Alba$^1$}
\address{$^1$
 Max-Planck-Institut f\"{u}r Physik komplexer Systeme,
  N\"{o}thnitzer Stra{\ss}e 38, D-01187 Dresden, Germany}

\author{Masudul Haque$^2$}
\address{$^2$
 Max-Planck-Institut f\"{u}r Physik komplexer Systeme,
  N\"{o}thnitzer Stra{\ss}e 38, D-01187 Dresden, Germany}

\author{Andreas~M.~L\"auchli$^3$}
\address{$^3$
 Institute f\"ur Theoretische Physik, Universit\"at 
Innsbruck, A-6020 Innsbruck, Austria}

\date{\today}

\begin{abstract} 

We study the entanglement spectrum (ES) of a finite XXZ spin-$1\over 2$ chain
in the limit $\Delta\to-1^+$ for both open and periodic boundary conditions.
At $\Delta=-1$ (ferromagnetic point) the model is equivalent to the 
Heisenberg ferromagnet and its degenerate ground state 
manifold is the $SU(2)$ multiplet with maximal total spin.  Any state in 
this so-called ``symmetric sector'' is an equal weight
superposition of all possible spin configurations.  In the gapless phase at
$\Delta>-1$ this property is progressively lost as one moves away from the
$\Delta=-1$ point.  We investigate how the ES obtained from the states in this
manifold reflects this change, using exact diagonalization and Bethe ansatz
calculations.  We find that in the limit $\Delta\to-1^+$ most of the ES levels
show divergent behavior.  Moreover, while at $\Delta=-1$ the ES contains no
information about the boundaries, for $\Delta>-1$ it depends dramatically on
the choice of boundary conditions.  For both open and periodic boundary
conditions the ES exhibits an elegant multiplicity structure for which we
conjecture a combinatorial formula.  We also study the entanglement
eigenfunctions, i.e. the eigenfunctions of the reduced density matrix.  We
find that the eigenfunctions corresponding to the non diverging levels mimic
the behavior of the state wavefunction, whereas the others show intriguing
polynomial structures.  Finally we analyze the distribution of the ES levels
as the system is detuned away from $\Delta=-1$.

\end{abstract}

\maketitle

\section{Introduction}
\label{intro}

Entanglement is one of the most fascinating concepts in quantum mechanics and
the application of entanglement related ideas to condensed matter in recent
years has boosted a huge interdisciplinary
effort \cite{AmicoFazioOsterlohVedral_RMP08}. In particular there has been a
growing interest in studying the properties of the so
called \emph{entanglement spectrum}
(ES) \cite{Peschel_papers,topological_ES_various, Alba_PRL2012,Nienhuis-09, Poilblanc-10,
Schliemann_2011, Santos-11, Cirac-11, cal-lev-2008}.  
The original motivation for the study of the ES was that the ES lies at the heart 
of the density matrix renormalization group (DMRG) algorithm \cite{Peschel_papers}.  
More recently, the ES has attracted a lot of attention
because of its relation to low-energy boundary modes in
topological phases \cite{topological_ES_various}.

Considering a bipartition of a system into two parts $A$ and $B$, the
 entanglement spectrum (ES), $\{\xi_i\}$, is defined in terms of the Schmidt
 decomposition
\begin{equation}  \label{eq:schmidt_ES_defn}
|\psi\rangle=\sum_i e^{-\xi_i/2}|\psi_i^A\rangle
\otimes |\psi_i^B\rangle. 
\end{equation}
where $\ket{\psi}$ is the state of the system, and
$|\psi_i^A\rangle$($|\psi_i^B\rangle$) form an orthonormal basis for the
subsystem $A$ ($B$).  The ES $\{\xi_i=-\log\lambda_i\}$ can also be thought of
in terms of the eigenvalues $\{\lambda_i\}$ of the reduced density matrix
$\rho_A$ obtained after tracing out the $B$ part of the system density matrix
$\ket{\psi}\bra{\psi}$.

In this work we study the ES of the finite spin-$\frac{1}{2}$ anisotropic
Heisenberg (XXZ) chain in the vicinity of the ferromagnetic point $\Delta=-1$.
(Here $\Delta$ is the anisotropy parameter.)  In particular we focus on the
transformation from the gapless phase at $-1<\Delta\le 1$, described by a
$c=1$ conformal field theory (CFT), to the $\Delta=-1$ point.  In the gapless
phase (except for the free fermion point $\Delta=0$ where many results are
available \cite{Peschel_papers}) only the distribution of the ES levels is
known from conformal invariance \cite{cal-lev-2008}.  On the other hand at
$\Delta=-1$ the system is not conformal invariant and a dramatic change
occurs. This is reflected in a peculiar behavior of usual entanglement 
related quantities such as the Renyi entropies and the concurrence~
\cite{ercolessi-2011,ercolessi-2012,colomo-2009}. In particular, the Renyi entropies in the 
vicinity of $\Delta=-1$ show signature of an essential singularity~\cite{ercolessi-2011,
ercolessi-2012}. 
The purpose of this work is to understand how this change in the behavior of 
the system is reflected at the level of the ES. 

The $\Delta=-1$ point is unitarily related to the $SU(2)$-invariant Heisenberg
ferromagnetic chain, and hence the ground state is the degenerate $SU(2)$
multiplet corresponding to the highest total spin $S_T$.  At $\Delta=-1$ the
states of the multiplet are flat superpositions of all the possible spin
configurations compatible with total magnetization, $S^z_T$, i.e., the
particles are fully delocalized.  We therefore refer to the ground state
multiplet as the symmetric multiplet.  In other terms at $\Delta=-1$ the
ground state shows a ``mean field'' structure, i.e., the system has no notion
of distance.  This implies for example that there is no difference between
open and periodic boundary conditions.

While there are well-known many-body model systems lacking notions of
distance, such as the Lipkin-Meshkov-Glick model~\cite{
lip-mesh-1965,mesh-lip-1965,glick-lip-1965,lat-vid-2005} and the
Richardson model~\cite{ric-63}, these are generally mean-field like at the
level of the Hamiltonian.  The specialty of the present case is that the XXZ
interaction is local, and the symmetric structure appears only at a single
value ($\Delta=-1$) of the anisotropy parameter.  Varying $\Delta$ is a
natural way to tune away from the symmetric structure.  In this Article, we
show how this change from the symmetric to the more usual situation is
manifested in the ES.

Although the usual interest is in entanglement properties of the ground state,
in the vicinity of $\Delta=-1$ it is natural to study the entire set of states
which become degenerate at that point.  The ES properties we report are found
to be common to the entire symmetric sector.  For simplicity of demonstration
or explicit calculations, we find it  convenient to sometimes focus on simpler
large-$S^z_T$ members of this manifold (which in the energy spectrum are near 
the top of this group of states for $\Delta>-1$) in addition to the $S^z_T=0$ 
ground state.

At $\Delta=-1$, the symmetric structure fully determines the
ES \cite{pop-sal-2005,pop-sal-2010,cas-doy-2011, cas-doy-2011-frac}. However,
at $\Delta>-1$ the ES is not restricted, and in the limit $\Delta\to-1^+$ 
most of the ES levels diverge (i.e. most of the eigenvalues of
$\rho_A$ vanish).  The divergent levels correspond to components in the ground
state wavefunction at $\Delta>-1$ that do not reflect the symmetric structure
present at $\Delta=-1$.  Moreover, while at $\Delta=-1$ the ES does not
contain any information about the boundary conditions or the geometry of the
partitioning between $A$ and $B$, for $\Delta>-1$ the ES for periodic and open
boundary conditions display striking differences.  For open boundary
conditions the ES is less dense (some of the ES levels are missing).  This is
related to there being two boundaries between $A$ and $B$ for periodic
boundary conditions (and only one for open).  This difference is also present
in the behavior of the entanglement entropy in the gapless phase (as
$1/3\log\ell$ for periodic and as $1/6\log\ell$ for open boundary conditions
with $\ell$ the length of subsystem $A$).  We also find that the multiplicity
count of the ES in the vicinity of $\Delta=-1$ (numbers of ES levels diverging
at different rates) seems to be described by an elegant combinatorial formula
which we present as a conjecture.

We also analyze the entanglement eigenfunctions, i.e. the eigenstates of
$\rho_A$.  While the eigenfunctions corresponding to the lower (non-diverging)
ES levels mimic the behavior of the ground state wavefunction (i.e. they
exhibit a symmetric structure in the limit $\Delta\to-1^+$), the others show
richer structures and in some cases their leading behavior (in $\epsilon
\equiv\Delta+1$) can be given in terms of known polynomial functions.

Finally we investigate the distribution of the ES levels in the vicinity of
the ferromagnetic point.  Although in the thermodynamic limit this is expected
to be a universal function, for finite chains the ferromagnetic point
introduces significant corrections in the whole region $\Delta<0$.

For our analysis we take advantage of the fact that the XXZ is an integrable
model and all its eigenstates can be in principle calculated using Bethe
ansatz.  We also use exact numerical diagonalization.  Since ES levels
are determined often up to values of $\sim100$, the reduced density
matrix eigenvalues $\lambda_i=e^{-\xi_i}$ need to be calculated with very
high precision.  Therefore, as in Ref.~\cite{Alba_PRL2012}, we use arbitrary
precision numerics.


In section \ref{XXZ_limit} we discuss the spin-$\frac{1}{2}$ $XXZ$ chain
around $\Delta=-1$, showing the ground state manifold and how the
wavefunctions of this symmetric sector behave at $\Delta\to-1^+$. In
section \ref{THE_ES} we present the main features of the ES, reviewing the
exact structure known at $\Delta=-1$ and then presenting numerical
diagonalization data for $\Delta=-1+\epsilon$.  In section~\ref{pert_ES} we
treat the ES perturbatively in the distance $\Delta+1$ from the ferromagnetic
point, using the Bethe ansatz.  The combinatorial aspects of the ES are also
addressed and characterized.  Section~\ref{eig_fct} is devoted to the study of
the entanglement eigenfunctions.  In section~\ref{away} we investigate how the
distribution of the ES levels changes as the system is detuned away from the
ferromagnetic point.  Details of the Bethe ansatz calculation appear in the
Appendix.

\section{The XXZ in the limit $\Delta\to-1^+$: the symmetric 
multiplet, energy spectrum and wavefunctions}
\label{XXZ_limit}

The spin-$\frac{1}{2}$ $XXZ$ open chain of $L$ sites is defined by the
Hamiltonian
\begin{equation}
{\mathcal H} ~=~ \frac{1}{2}\sum\limits_{i=1}^{L-1} 
(S^+_iS^-_{i+1}+ S^i_iS^+_{i+1}) ~+~
\Delta\sum\limits_{i=1}^{L-1} S^z_iS^z_{i+1}
\end{equation}
and the periodic chain is obtained by adding terms connecting sites $L$ and 1.
At $\Delta=-1$ the model is related to the ferromagnetic Heisenberg model
($-\sum_i \vec{S}_{i}\cdot\vec{S}_{i+1}$) by the local unitary transformation
$S^{x,y}\to -S^{x,y}$ applied on every second site of the
chain~\footnote{Since this transformation is a product of local unitary
transformations, it does not affect the entanglement spectrum.}.
The ground state of the Heisenberg ferromagnet is in the $SU(2)$ multiplet
with highest total spin $S_T$ (symmetric multiplet), i.e. the one containing
the fully polarized state $\ket{F}\equiv
\ket{\u\u\u\cdots}$. 
The $L+1$ states of the degenerate multiplet are obtained from $\ket{F}$ by
successive applications of the lowering operator $S^-_T\equiv\sum_iS^-_i$.  It
is conventional to refer to $|F\rangle$ as the ``vacuum'' and the overturned
spins as `particles'.  The states $(S_T^-)^M|F\rangle$ are equal weight
superpositions of all the possible spin configurations with $M$ down spins
(particles).  The magnetization is $S^z_T=L/2-M$ in terms of the particle
content $M$; either ($S^z_T$ or $M$) is a good quantum number for all $\Delta$
and labels the states of the multiplet.

\begin{figure}
\begin{center}
\includegraphics[width=.8\textwidth]{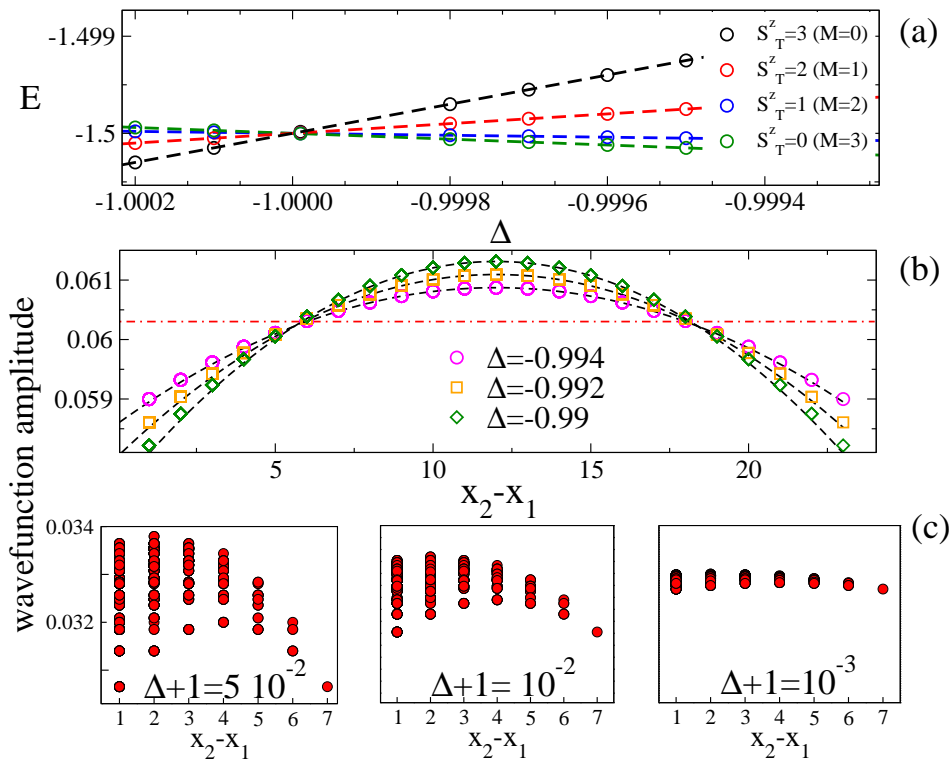}
\end{center}
\caption{ ({\bf a}) Energy spectrum around $\Delta=-1$ 
 of the XXZ spin chain with $L=6$ and periodic boundary conditions. We show
 the energies of all the chain eigenstates within the symmetric multiplet and
 $S^z_T\ge 0$.  The energy at $\Delta=-1$ is $E=-L/4$.  The dashed lines are
 from Eq.~\eref{ener_ser}.  ({\bf b}) The wavefunction of the state with only
 two particles ($M=2$, $S^z_T=L/2-M$) for a chain with $L=24$ and several
 values of $\Delta$. We plot the wavefunction components versus the
 interparticle distance $x_2-x_1$ ($x_2>x_1$).  The points are exact
 diagonalization data while the dashed line is the Bethe ansatz result,
 Eq.~\eref{two_part_bethe}).  The dashed-dotted line is the asymptotic
 ($\Delta=-1$) value, $\sqrt{2/[L(L-1)]}$.  ({\bf c}) The wavefunction of the
 state with $M=L/2$ ($L=12$) particles (ground state at $\Delta>-1$) plotted
 versus the relative distance of the first two particles.  Data are obtained
 using Eq.~\eref{many_part_bethe}.  We show the wavefunction for three
 different values of $\Delta+1$, namely $\Delta+1=5\cdot 10^{-2},
 10^{-2},10^{-3}$.  At each fixed $x_2-x_1$ the different points correspond to
 the possible configurations of the other $L/2-2$ particles.  }
\label{two_part_gs}
\end{figure}

In Figure \ref{two_part_gs} ({\bf a}) we show the energies of the states which
are connected to the symmetric multiplet at $\Delta=-1$, as a function of
$\Delta+1$.  The energy levels in the vicinity of $\Delta=-1$ are
\begin{equation}
\label{ener_ser}
E=-\frac{L}{4}+\Bigg[\frac{L}{4}-\frac{1}{L-1}
\Big(\frac{L}{2}-S^z_T\Big)\Big(\frac{L}{2}+S^z_T\Big)
\Bigg](\Delta+1)   \, .
\end{equation}

The ground state at $\Delta>-1$ is the one with $S^z_T=0$, while the other
states of the multiplet are excited states.  In Figure \ref{two_part_gs} ({\bf
b}) we plot the wavefunction of the eigenstate with two particles (which is
part of the symmetric multiplet but not the ground state) showing wavefunction
components versus the relative position ($x_2-x_1$) of the two particles.  While at the
point $\Delta=-1$ the wavefunction is completely flat (the two particles are
completely delocalized), for $\Delta>-1$ this property is progressively lost
farther away from $\Delta=-1$.

This behavior is true for any state in the symmetric multiplet, although more
complicated to display.  For example in Fig.~\ref{two_part_gs} ({\bf c}) we
show wavefunction components of the ground state ($S^z_T=0$) for several
values of $\Delta$.  Now there are multiple components for each value of
$x_2-x_1$.  It is clear, however, that in the limit $\Delta\to-1$ the
wavefunction becomes flatter and eventually becomes an equal weight
superposition.

\section{Entanglement spectrum (ES) of the symmetric multiplet: general features}
\label{THE_ES}

In this section we describe how the structure of the ground
 state in the vicinity of $\Delta=-1$  
determines the behavior of the ES. To calculate
 the ES we consider a bipartition of the chain 
into two parts $A$ and $B$ (of lengths $L_A$ and 
$L_B$ respectively). The wavefunction 
$\ket{\psi}$ of the chain, given two 
orthonormal bases for the subsystems 
$|\varphi_i^A\rangle,|\varphi_j^B\rangle$ can be written
as 

\begin{equation}
\label{schmidt_matrix}
\ket{\psi}=\sum\limits_{i,j}{\bf M}_{ij}|
{\varphi^A_i}\rangle\otimes|{\varphi^B_j}\rangle
\end{equation}

The matrix ${\bf M}_{ij}$ is a central object in the study of ES.  The reduced
density matrix eigenvalues $\lambda_i=e^{-\xi_i}$
[Eq.~\eref{eq:schmidt_ES_defn}] are the square of the Schmidt (singular)
values of ${\bf M}$.  Since the block magnetization $S^z_A$ is a good quantum
number for the matrix ${\bf M}$ we can label the ES levels with $S^z_A$.

\subsection{The point $\Delta=-1$}
\label{ferro_ES}

At $\Delta=-1$ the flat structure of the states in the symmetric sector, makes
it possible to construct the matrix ${\bf M}$ explicitly and to calculate its
singular values $\sqrt{\lambda}_i$. Any eigenstate in the symmetric multiplet
is identified given the length of the chain $L$ and the magnetization $S^z_T$
(particle content).  For each of these eigenstates there is one ES level per
block magnetization sector $S^z_A$ which is given by \cite{pop-sal-2005}

\begin{equation}
\label{pop-sal-2005}
\xi^{\{L,L_A\}}_{\{S^z_T,S^z_A\}}=-\log\left[
\frac{{L_A\choose \frac{L_A}{2}-S^z_A}
{L-L_A\choose\frac{L-L_A}{2}-S^z_T+S^z_A}}
{{L\choose\frac{L}{2}-S^z_T}}\right]
\end{equation}

As a consequence of the delocalized structure of the 
states in the symmetric multiplet, Eq.~\eref{pop-sal-2005} 
holds for both periodic and  open boundary conditions.

After restricting to $S^z_T=0$,  that is the 
ground state at $\Delta>-1$, and taking the 
thermodynamic limit $L\to\infty$ one obtains

\begin{equation}
\label{cas-doy-2011}
\xi^{\{L_A\}}_{\{S^z_A\}}=-\log\left[\frac{1}{2^{L_A}}
{L_A\choose\frac{L_A}{2}-S^z_A}\right]
\end{equation}

which has also been obtained in~\cite{cas-doy-2011}.

\begin{figure}
\begin{center}
\includegraphics[width=.9\textwidth]{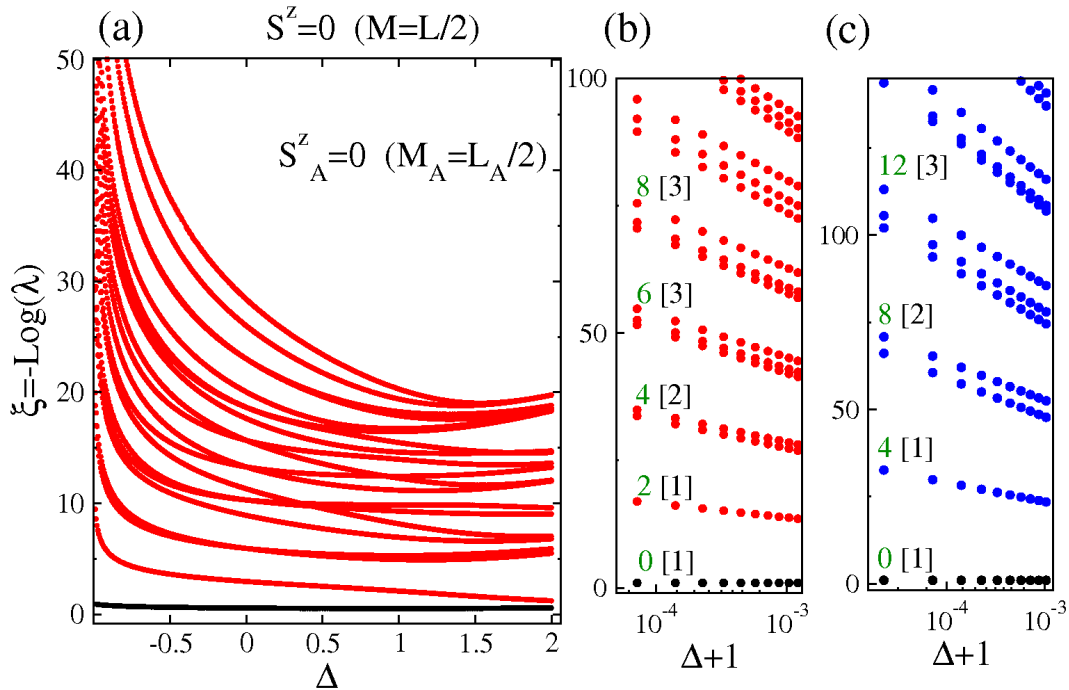}
\end{center}
\caption{Entanglement spectrum from exact diagonalization ($L=18$, $L_A=6$),
  plotted against $\Delta$.  We show only the $S^z_A=0$ sector.  In the limit
  $\Delta\to -1$ the lowest ES level (in black) converges to the finite value
  given by \eref{pop-sal-2005}, while all the other levels diverge.  ({\bf a})
  Periodic boundary conditions; wide range of $\Delta$.  ({\bf b}) Periodic
  boundary conditions; zooming around $\Delta=-1$.  Log-scale on horizontal
  axis highlights the logarithmic divergence as $-\alpha\log(\Delta+1)+\beta$
  of the levels.  There are multiple ES levels for each value of $\alpha$.  We
  use the notation $\alpha [m_{per}]$ where $m_{per}$ is the multiplicity of
  the set of levels with same exponent $\alpha$.  ({\bf c}) The same as in
  ({\bf b}) but for a chain with open boundary conditions.  }
\label{spec_intro}
\end{figure}

\subsection{The ES in the limit $\Delta\to -1^+$: exact 
diagonalization results}
\label{ES_limit}

In Fig.~\ref{spec_intro} ({\bf a}) we show the ES in the vicinity of
$\Delta=-1$ (i.e. $\Delta+1\equiv\epsilon=0$) for a block with $L_A=6$ in a
chain with periodic boundary conditions and $L=18$ sites. For clarity we show
only the sector of the ES with $S^z_A=0$, since the result in other $S^z_A$
sectors is qualitatively similar.

The ES levels show singular behavior. In the limit $\Delta\to -1^+$ there is
only one level converging to a finite value given by Eq.~\eref{pop-sal-2005},
while all the others diverge.  The diverging behavior in the ES signals that
in the limit $\Delta=-1^+$ the wavefunction has to be a symmetric
superposition of all allowed configurations.  The components in the total
wavefunction which reflect this symmetry give rise to a non diverging level,
while the others contribute to the diverging ones.  In Section \ref{eig_fct}
we will show how this distinction is manifested in the structure of
entanglement eigenfunctions corresponding to diverging and non-diverging ES
levels.

In Fig.~\ref{spec_intro} ({\bf b}) we demonstrate that the behavior of the
levels in the vicinity of $\Delta=-1$ is given by
$-\alpha\log(\Delta+1)+\beta$, $\alpha$ being an integer. The slopes $\alpha$,
going from bottom to top in the ES, are given by the sequence $0,2,4,6,...$
where $0$ refers of course to the lowest level.  A diverging level in the ES
as $-\alpha\log (\Delta+1)$ corresponds to an eigenvalue of the reduced
density matrix which vanishes as $(\Delta+1)^\alpha$.  Focusing on the number
of levels diverging with the same slope, an intriguing multiplicity sequence
appears as higher levels in the ES are considered.  In Fig.~\ref{spec_intro}
({\bf b}) we use the notation $\alpha[m_{per}]$, $m_{per}$ being the number of
levels with slope $\alpha$. We observed that $m_{per}(\alpha)$ is given by the
sequence $m_{per}(\alpha)=1,1, 2,3,\dots$.

In Fig.~\ref{spec_intro} ({\bf c}) we show the same plot but for a chain with
open boundary conditions ($L=17$).  The ES shows the same qualitative
behavior, although now the slopes are $\alpha=0,4,8,12\dots$.  The
multiplicity structure apparently is described by the sequence
$m_{open}(\alpha)=m_{per}(\alpha/2)$.

In a generic sector with $S^z_A \ne 0$, the slope sequence (the $\alpha$'s)
remains the same ($\alpha=0,2,4,\dots$ for periodic boundary conditions and
$\alpha=0,4,8,\dots$ for open boundary conditions), but the multiplicity
structure is different.

\begin{figure}
\begin{center}
\includegraphics[width=.9\textwidth]{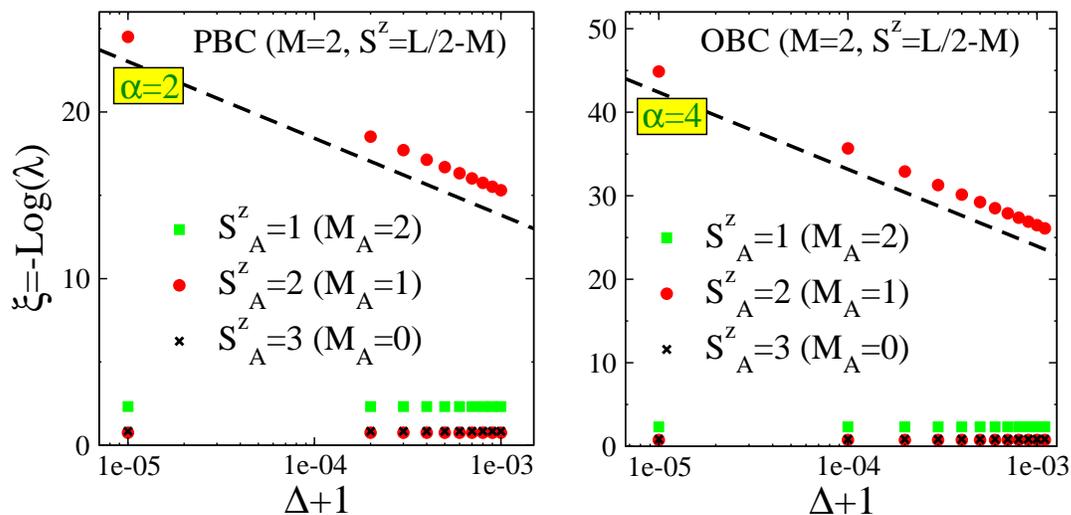}
\end{center}
\caption{ ES for the member of the symmetric multiplet containing
only two particles ($M=2$), plotted versus $\Delta+1$ for periodic ({\bf PBC})
and open ({\bf OBC}) boundary conditions.  The block size is $L_A=6$.  $S^z_A$
($M_A$) is the magnetization (particle content) of block $A$.  The behavior of
the diverging ES levels is $\xi=-\alpha\log(\Delta+1)+\beta$. The dashed lines
are guides to the eye.  }
\label{two_part_spec}
\end{figure}

The outlined scenario depends only on the symmetric structure of the states in
the limit $\Delta\to-1^+$, thus the ES obtained from any state within the
symmetric multiplet is expected to show similar features.  This observation is
supported in Fig.~\ref{two_part_spec} where we show the full ES obtained from
the state in the symmetric multiplet with only two particles (which is not the
ground state for $\Delta>-1$, see Fig.~\ref{two_part_gs}).  We show the ES for
both periodic (left) and open boundary conditions (right). In both cases there
are only three possible sectors $S^z_A$ determined by the number of ways of
distributing the two particles in $A$ and $B$.  Nonetheless, the main features
observed in Fig.~\ref{spec_intro} are present: (a) for each value of $S^z_A$
there is one ES level converging to a finite value in the limit $\Delta\to
-1^+$ and this value is given by Eq.~\eref{pop-sal-2005}; (b) the slope of the
only divergent level is $2$ for periodic boundary conditions and $4$ for open,
in agreement with the sequences shown in Fig.~\ref{spec_intro}.

\section{The ES at $\Delta\to -1^+$: 
perturbative and combinatorial structure}
\label{pert_ES}

The structure of the ES outlined in the previous sections can be
quantitatively understood by expanding the state wavefunction for small
$\epsilon\equiv\Delta+1$ and then calculating the ES. For the XXZ we can
exploit the fact that the model is integrable and use the exact expression of
the eigenstate wavefunctions (see~\ref{bethe_pert} for details).

For this description, we choose the orthonormal basis
$\{|\varphi^A_i\rangle\}=\{\ket{x_1,x_2,
\cdots,x_{M_A}}\}$ for the $A$ block,  where $|x_1,x_2,
\cdots,x_{M_A}\rangle$ are the block configurations 
with particles at positions $x_1,x_2,\cdots,x_{M_A}$.  Similarly for the $B$
subsystem we have $\{|\varphi^B_i\rangle\}=\{\ket{x_{M_A+1},x_{M_A+2},
\cdots,x_{M}}\}$.  

With this notation the elements of the ${\bf M}$ matrix
[Eq.~\eref{schmidt_matrix}] are
\begin{equation}
\label{many_part_M}
{\bf M}^{[x_1,x_2,\cdots,x_{M_A}]}
_{[x_{M_A+1},\cdots,x_M]}\equiv a(x_1,x_2,\cdots,x_{M_A},
x_{M_A+1},\cdots,x_M)
\end{equation}
where $[x_1,x_2,\cdots, x_{M_A}]$ and $[x_{M_A+1},\cdots,x_M]$ act as row or
column indices.  Here $a(x_1,x_2,\cdots,x_{M_A},x_{M_A+1},\cdots,x_M)$ is the
amplitude of the chain configuration with particles at positions
$x_1,x_2,\cdots,x_{M_A}, x_{M_A+1},\cdots,x_M$.  It turns out that the first
order in $\Delta+1$ for the amplitude $a$ in \eref{many_part_M} can be
obtained analytically for arbitrary $M$ and system sizes $L$ and the resulting
ES calculated.

In the three subsections below, we first treat the simplest nontrivial member
of the symmetric multiplet (subsection \ref{first_ES_tp}), namely the one with
only two particles ($M=2$), and provide the first order expansion in
$\epsilon\equiv\Delta+1$.  In \ref{comb_ES}, we discuss $M>2$ and higher
orders in $\epsilon$, providing a conjecture for the multiplicity sequence of
the ES at different orders.  In \ref{sec_ES_open} we comment on the
open-boundary case.

\subsection{ The sector with two 
particles ($M=2$): first order in $\Delta+1$}
\label{first_ES_tp}

In the sector with two particles in the full chain wavefunction, 
using the expansion \eref{two_part_bethe} 
and \eref{many_part_M} the ${\bf M}$ matrix can be 
written explicitly.  The possible values of $M_A$ are $M_A=0,1,2$.  Moreover
the blocks of ${\bf M}$ corresponding to $M_A=0,2$ are $1\times 1$ (i.e. just
numbers).  In Fig.~\ref{two_part_spec} ({\bf PBC}) these would give the two ES
levels with $S^z_A=1,3$, which are not diverging in the limit $\Delta\to
-1^+$.  For these two levels the first order contribution to the wavefunction
gives a first order renormalization of the result~\eref{pop-sal-2005}. After
restricting to the case $L_A=L_B=L/2$ to simplify the expressions that would
be otherwise very cumbersome, the ES level in the sector with $M_A=2$ is
written as
\begin{equation}
\label{triv}
-2\log\left[\frac{1}{2}\sqrt{\frac{L-2}{L-1}}\Big(1-
\frac{L(L-2)}{24(L-1)}(\Delta+1)\Big)\right]
\end{equation}

Since we chose $L_A=L_B$ the ES level with $M_A=0$ is the same.

More interesting is the sector with $M_A=1$ ($S^z_A=2$ in Fig.~\ref{two_part_spec}), 
in which case there is a divergent level.  The block in matrix ${\bf M}$ 
describing this sector reads
\begin{equation}
\label{two_part_m}
\fl{\bf M}=\sqrt{\frac{2}{L(L-1)}}\Big[(1+A_2\epsilon)
w^{(0)}\otimes w^{(0)}+\frac{\epsilon}{L-1}[w^{(0)}
\otimes w^{(2)}+0\leftrightarrow 2]-\frac{2\epsilon}
{L-1}w^{(1)}\otimes w^{(1)}\Big]
\end{equation}
where $\otimes$ is the Kronecker product and we defined the three vectors
$w^{(i)}_k\equiv k^i$ and $\epsilon\equiv\Delta+1$. The factor $A_2$ ensures
the normalization of the full chain wavefunction (see~\ref{bethe_pert}). ${\bf
M}$ is a symmetric matrix and quadratic polynomial as a function of the
coordinates of the two particles.  Since the vectors $w^{(i)}$ are linearly
independent, the rank of ${\bf M}$ is three. However, while two of the
singular values of~\eref{two_part_m} are ${\mathcal O}(1)$ and ${\mathcal
O}(\epsilon)$, the other one is ${\mathcal O}(\epsilon^2)$ and it is not
meaningful because the wavefunction is accurate only up to first order. The
third singular value is thus ``fake'' and the effective rank of
~\eref{two_part_m} is two.  (See Ref.~\cite{Alba_PRL2012} for a detailed
discussion of such spurious eigenvalues in perturbative ES calculations.)  One
ES level of~\eref{two_part_m} is given by
\begin{equation}
\label{first}
-2\log\left[\frac{1}{24\sqrt{2}}\frac{L^2(L+2)(L-2)}
{(L(L-1))^{\frac{3}{2}}}(\Delta+1)\right]
\end{equation}
which diverges in the limit $\Delta\to-1$ and corresponds to 
the divergent ES level in Fig.~ \ref{two_part_spec} 
({\bf PBC}). The non diverging ES level 
in this sector up to first order is given by
\begin{equation}
\label{low}
-2\log\left[\sqrt{\frac{L^2}{2L(L-1)}}+\frac{(L-2)[L(L-1)]^{\frac{3}{2}
}}{12\sqrt{2}(L-1)^3}(\Delta+1)\right]  \, .
\end{equation}
Note that the correction to the singular values, at first order in
$\epsilon\equiv(\Delta+1)$, is always of type ${\sim}L\epsilon$ for 
large $L$ [Eqs.~\eref{triv},\eref{first},\eref{low}].   Thus
$L\epsilon=L(\Delta+1)$ needs to be small for these expansions to be
meaningful.

\subsection{Many particles ($M>2$) and higher orders in 
$\Delta+1$: combinatorial structure of the ES}
\label{comb_ES}

In this section we discuss how the structure of the ES for any state within
the symmetric multiplet builds up as the particle number ($M$) of the
wavefunction increases and higher orders in $\Delta+1$ are taken into account.

In each $M_A$ sector the ES levels are organized in powers of
$\Delta+1$: the lowest ES level converges to a finite value in the limit
$\Delta\to -1$, while the higher levels diverge as
$-\alpha\log(\Delta+1)+\beta$ with $\alpha=2,4,6,...$.  as observed in
Fig.~\ref{spec_intro}.

Within a given $S^z_A$ sector and for each value of the slope $\alpha$ 
there is more than one level, i.e. a non trivial multiplicity 
structure arises as higher levels in the ES are considered.

\begin{figure}
\begin{center}
\includegraphics[width=1\textwidth]{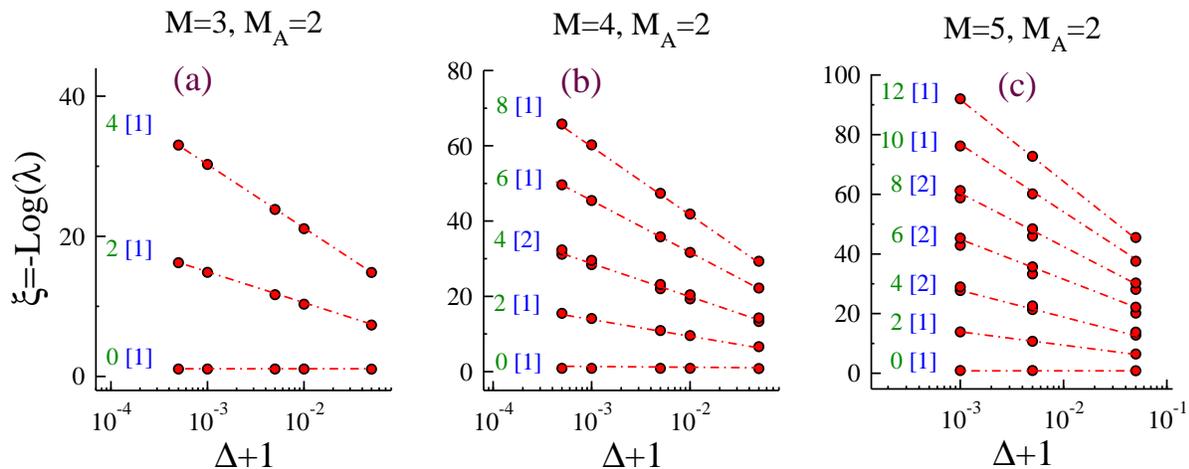}
\end{center}
\caption{ ES for a chain of $L=18$ sites, $L_A=8$,
 periodic boundary conditions and several values of $\Delta$ (arbitrary
 precision exact diagonalization data). $M$ ($M_A$) is the number of particles
 in the full chain ($A$ block).  Logarithmic scale on the horizontal axis
 highlights the divergent behavior ($\xi=-\alpha \log(\Delta+1) +\beta$ with
 $\alpha=0,2,4,...$).  We use the same notation $\alpha[m_{per}]$ as in
 Fig.~\ref{spec_intro} for divergence exponent and multiplicity.  }
\label{many_part_comb}
\end{figure}

It  is natural to investigate how the multiplicity sequence depends
on the particle content ($M$). To this purpose in 
Fig.~\ref{many_part_comb} we provide examples of multiplicity sequences in
the ES for $M_A=2$ and $M=3,4,5$ (we consider a chain with $L=18$ and $L_A=8$).
We note that the sequences can be obtained as particular cases from the expansion of 
the so called q-deformed binomial which is defined as:
\begin{equation}
\label{conj}
C_q(r,s)={r\choose s}_q\equiv\frac{\prod\limits_
{k=r}^{r-s+1}(1-q^{k})}{\prod\limits_
{k=1}^{r} (1-q^k)}
\end{equation}
For example, if we fix $M=r=5,M_A=s=2$ we obtain $C_q(5,2)=1+q+2q^2+2q^3+2q^4+
q^5+q^6$. The sequence $1,1,2,2,2,1,1$ given by the coefficients of the different monomials
matches the one shown in Fig.~\ref{many_part_comb} ({\bf c}).  It can be
checked that $C_q(4,2),C_q(3,2)$ give the other sequences in
Fig.~\ref{many_part_comb}. We also checked that the result holds for other
values of $M,M_A$. Thus we conjecture that for all the states within the
symmetric multiplet the multiplicity of the ES level diverging as $-\alpha\log
(\Delta+1)$ is given by the coefficient of $q^{\alpha/2}$ in the expansion of
the q-deformed binomial~\eref{conj} with $r=M, s=M_A$. 
This result holds provided that $M<L_A$~\footnote{ 
Here we are considering the case $L_A\le L_B$ otherwise the condition
is $M< min(L_A,L_B)$. However, since the reduced density matrices
 of the two subsystems are isospectral we can always choose subsystem $A$
 such that $L_A\le L_B$.}.  We observe that for $L_A\le M\le L_B$ the multiplicity sequence 
is given by  \eref{conj} with $r=L_A,s=M_A$, while if $M>L_B$ it is $r=L_A-(M-L_B)$ 
$s=M_A-(M-L_B)$.

\subsection{Open boundary conditions}
\label{sec_ES_open}

The XXZ chain with open boundary conditions can also be treated using the
Bethe ansatz (details in~\ref{bethe_pert}).  The first order expansion of the
matrix ${\bf M}$ is

\begin{equation}
\fl
\label{open_many_part_M}
{\bf M}^{[x_1,x_2,\cdots,x_{M_A}]}
_{[x_{M_A+1},\cdots,x_M]}= \mathbf{1}+
\epsilon\big[{\mathcal M}_1(x_1,x_2,\cdots,x_{M_A})+
{\mathcal M}_2(x_{M_A+1},\cdots,x_M)\big] 
\end{equation}

where ${\mathcal M}_1$ and ${\mathcal M}_2$ 
are polynomial functions that can be obtained from~\eref{open_bethe_state}. 
We stress that \eref{open_many_part_M} does not contain 
any cross terms of the form $x_ix_j$ with
$x_i\in A$ and $x_j\in B$ (note the difference with
 the result~\eref{two_part_m} for periodic boundary conditions). 
This means that  at first order in $\epsilon$ 
there is no entanglement between $A$ and $B$. 

\begin{figure}
\begin{center}
\includegraphics[width=1\textwidth]{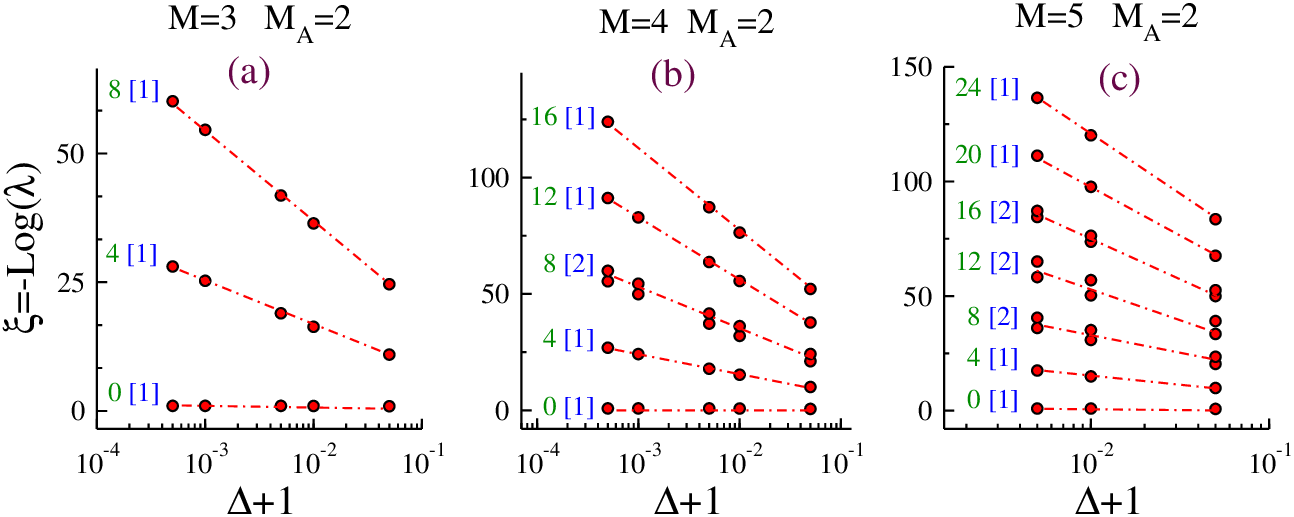}
\end{center}
\caption{ 
ES for a chain of $L=18$ sites, $L_A=8$,
 periodic boundary conditions and several values of $\Delta$ (arbitrary
 precision exact diagonalization data). $M$ ($M_A$) is the number of particles
 in the full chain ($A$ block).  Logarithmic scale on the horizontal axis
 highlights the divergent behavior ($\xi=-\alpha \log(\Delta+1) +\beta$ with
 $\alpha=0,4,8,...$).  We use the same notation $\alpha[m_{per}]$ as in
 previous figures \ref{spec_intro} and \ref{many_part_comb} for divergence
 exponent and multiplicity. 
 }
\label{open_many_part_comb}
\end{figure}

Despite the different values of $\alpha$ the multiplicity
structure in the ES shows similarities with the case of periodic
boundary conditions. In Fig.~\ref{open_many_part_comb} 
we show the ES levels in the sector 
with $M_A=2$ and total number of particles 
$M=3,4,5$. As already observed in Fig.~\ref{spec_intro}
all the ES levels with $\alpha=2k$ and $k$ odd
are missing. Nonetheless the multiplicity sequence 
$m_{open}(\alpha)$ is simply  related to the one observed 
for periodic boundary conditions as $m_{open}(\alpha)=
m_{per}(\alpha/2)$ with $m_{per}$ given in terms of~\eref{conj}.

\section{The entanglement eigenfunctions
at $\Delta\to-1^+$}
\label{eig_fct}

In this section we examine the entanglement eigenfunctions (eigenvectors of
reduced density matrix).   We show contrasting behavior for the eigenfunctions
corresponding to the ES levels which diverge in the $\Delta\to-1^+$ limit 
and those that do not diverge.  

We discuss the case of a chain with periodic boundary conditions.  We provide
information about structure of the entanglement eigenfunctions in two ways.
First, up to first order in $\Delta+1$ the Bethe ansatz is tractable for
calculating the entanglement eigenfunctions  for arbitrary $M$,
$M_A$, $L$ (Section \ref{sec_eigfunc_1storder}).  Second, in the $M_A=1$
sector, numerical data reveals a polynomial sequence for the forms of the
wavefunctions, which we express in terms of the discrete Chebyshev
polynomials (Section \ref{sec_entanglement_eigfiunc_chebyshev}).

\subsection{First order in $\Delta+1$} 
\label{sec_eigfunc_1storder}

\begin{figure}
\begin{center}
\includegraphics[width=.6\textwidth]{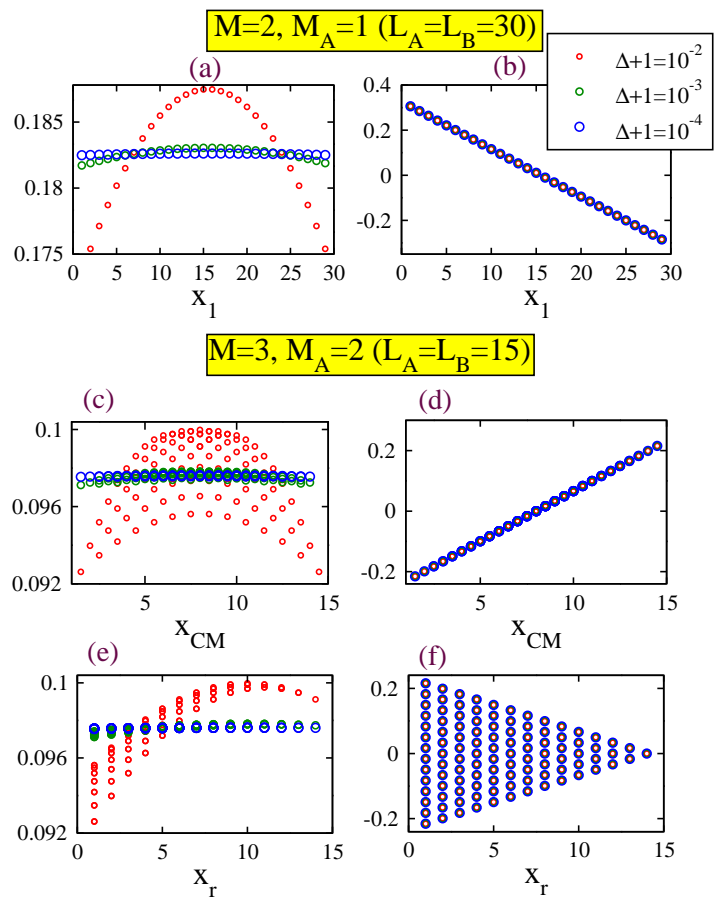}
\end{center}
\caption{ ({\bf a,b}) Entanglement eigenfunctions
 in the sector with only one particle in the block $A$ ($M_A=1$) and two
 particles in the full chain wavefunction ($M=2$) at different values of
 $\Delta$.  On the x-axis $x_1$ is the position of the particle inside block
 $A$. ({\bf a}) is the eigenfunction corresponding to the lowest non-diverging
 ES level while in ({\bf b}) we show the eigenfunction for the first diverging
 ES level (see Fig.~\ref{two_part_spec}).  ({\bf c,d,e,f}) Entanglement
 eigenfunctions in the sector with two particles in the block ($M_A=2$). In
 ({\bf c})({\bf e}) we show the entanglement eigenfunction for the lowest ES
 level versus $x_{CM}\equiv (x_1+x_2)/2$ and $x_r\equiv x_2-x_1$ where $x_1$
 $x_2$ are the position of the two particles. In ({\bf d})({\bf f}) we show
 the entanglement eigenfunction of the first divergent ES level.
 }
\label{two_part_eig}
\end{figure}

We start discussing the sector with $M=2$.  In this sector the entanglement
eigenfunctions reflect the polynomial structure of the matrix ${\bf M}$
(see~\eref{two_part_m}).  

In Fig.~\ref{two_part_eig} ({\bf a}) ({\bf b}) we show the ES eigenfunctions
for $M=2$ and $M_A=1$ obtained from the expansion~\eref{two_part_m}.  The
lowest ES level eigenfunction (({\bf a}) in Fig.~\ref{two_part_eig}) has a
parabolic dependence on $x_1$ (the position of the particle inside block $A$)
and as the ferromagnetic point is approached it becomes flat.
The non-diverging ES levels thus mimic the structure of the full chain
eigenstate.  The same behavior is present in the sectors with $M_A=0,2$, where
the ES has only one (non diverging) level.  In contrast, the eigenfunction of
the first diverging ES level shows linear behavior as a function of $x_1$
(Fig.~\ref{two_part_eig} ({\bf b})).

In Fig.~\ref{two_part_eig} ({\bf c})--({\bf f}) we show the entanglement
eigenfunctions for the case with two particles in the block $A$ ($M_A=2$) and
only one in $B$ ($M=3$). In ({\bf c}) and ({\bf d}) we plot the eigenfunction
components versus the ``center of mass'' coordinate $(x_1+x_2)/2$ of the two
particles in $A$. The entanglement eigenfunction of the lowest ES level ({\bf
  c}) exhibits the flattening behavior characteristic of the chain
wavefunction, while the eigenfunction in ({\bf d}) is again linear. The same
qualitative behavior is observed in the relative distance $x_2-x_1$ (see
Fig.~\ref{two_part_eig} ({\bf e}) ({\bf f})).

This result is general for all $M_A$, $M$ sectors --- the entanglement
eigenfunctions for the non-diverging ES levels flatten in the  limit
$\Delta\to-1^+$, while the eigenfunctions for the ES diverging as
$-\log(\Delta+1)$ are linear functions of the coordinates  $x_1,x_2,\dots,
x_{M_A}$.

\subsection{Entanglement eigenfunctions in the sector with
 $M_A=1$ and arbitrary $M$}
\label{sec_entanglement_eigfiunc_chebyshev}

\begin{figure}
\begin{center}
\includegraphics[width=.9\textwidth]{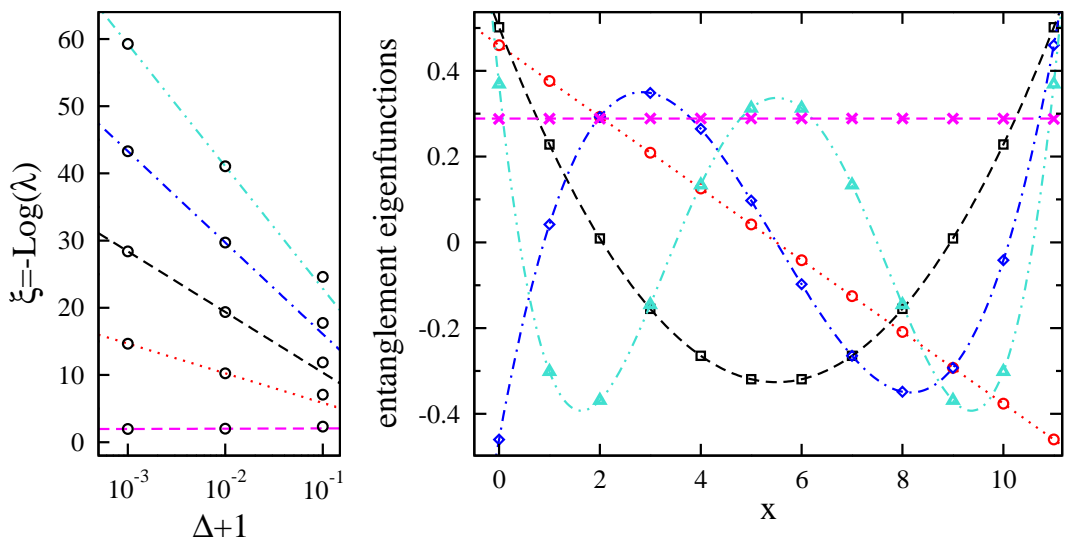}
\end{center}
\caption{ ES and entanglement eigenfunctions in the
 sector with $M_A=1$ (exact diagonalization data). ({\bf left}) The ES versus
 $\Delta+1$ for a block with $L_A=12$ embedded in a chain of size $L=24$. We
 show the ES levels in the sector with only one particle in the block
 ($M_A=1$) and five particles in the full chain ($M=5$).  The lines are fits
 to $-\alpha\log(\Delta+1)+\beta$.  ({\bf right}) For each level in the ES we
 show with the same color the components in the corresponding entanglement
 eigenfunction for $\Delta+1= 10^{-3}$.  Here $x$ is the position of the
 particle in block $A$. The lines are the polynomial behaviors given
 by Eq.~\eref{tche_eig}.  }
\label{five_part_poly}
\end{figure}

In the $M_A=1$ sector the number of ES levels is given by the total number of
particles $M$ [Eq.~\ref{conj}]. These levels are shown in
Fig.~\ref{five_part_poly} ({\bf left}) for $M_A=1$ and $M=5$. For each of the
ES levels we report ({\bf right}) the components of the corresponding
entanglement eigenfunction versus the position of the particle in block $A$.

We observe that in the limit $\Delta=-1^+$ the entanglement eigenfunction of
the level appearing at order $n$ (i.e. diverging as $-\alpha \log(\Delta+1)$
with $\alpha=2n$) is a polynomial of order $n$ (as a function of the position
of the particle in block $A$).  This is the only information needed in order
to know the exact functional form of the entanglement eigenfunctions.  The
polynomial can be fixed by imposing that the eigenfunctions are mutually
orthogonal and are normalized.

For instance the entanglement eigenfunction of the lowest
ES level  is the flat 
superposition $|\varphi_0\rangle=\sum_x c|x\rangle$, 
with $|x\rangle$ the configuration 
with the particle at position $x$ in 
block $A$. The coefficient $c$ is determined by 
imposing the normalization of the eigenfunction.
For  the first diverging
ES level  in Fig.~\ref{five_part_poly} (the 
red-dotted line) one has $|\varphi_1\rangle=
\sum_x(c^0+c^1 x)|x\rangle$. The coefficients $c^0,c^1$ are 
fixed by imposing that $\langle\varphi_0|\varphi_1\rangle=0$
and that $|\varphi_1\rangle$ is normalized. This gives

\begin{equation}
|\varphi_1\rangle=\sqrt{\frac{12}{L_A(L_A^2-1)}}
\sum_x\Big(x-\frac{L_A-1}{2}\Big)\ket{x}
\end{equation}
The next diverging ES level (the black-dashed line)
is instead

\begin{equation}
|\varphi_2\rangle=\sqrt{\frac{5}{L_A(L_A^4-5L_A^2+4)}}
\times\sum_x\Big(2+L_A^2+3L_A+6x^2-6(L_A-1)x\Big)\ket{x}
\end{equation}

The procedure allows to obtain the leading order 
of all the entanglement eigenfunctions in the sector $M_A=1$.
The set of polynomials obtained are known as the discrete Chebyshev 
polynomials $T_n$~\cite{sze-39} and are defined as

\begin{equation}
\label{tche_eig}
T_n(x)=n!{\mathcal D}^n {x\choose n} {x-L_A\choose n}
\end{equation}

where ${\mathcal D}$ is such that ${\mathcal D} f(x)
\equiv f(x+1)-f(x)$.

\section{Away from $\Delta= -1$: the ES distribution in the gapless
phase ($\Delta>-1$) }
\label{away}

\begin{figure}[h]
\begin{center}
\includegraphics[width=1\textwidth]{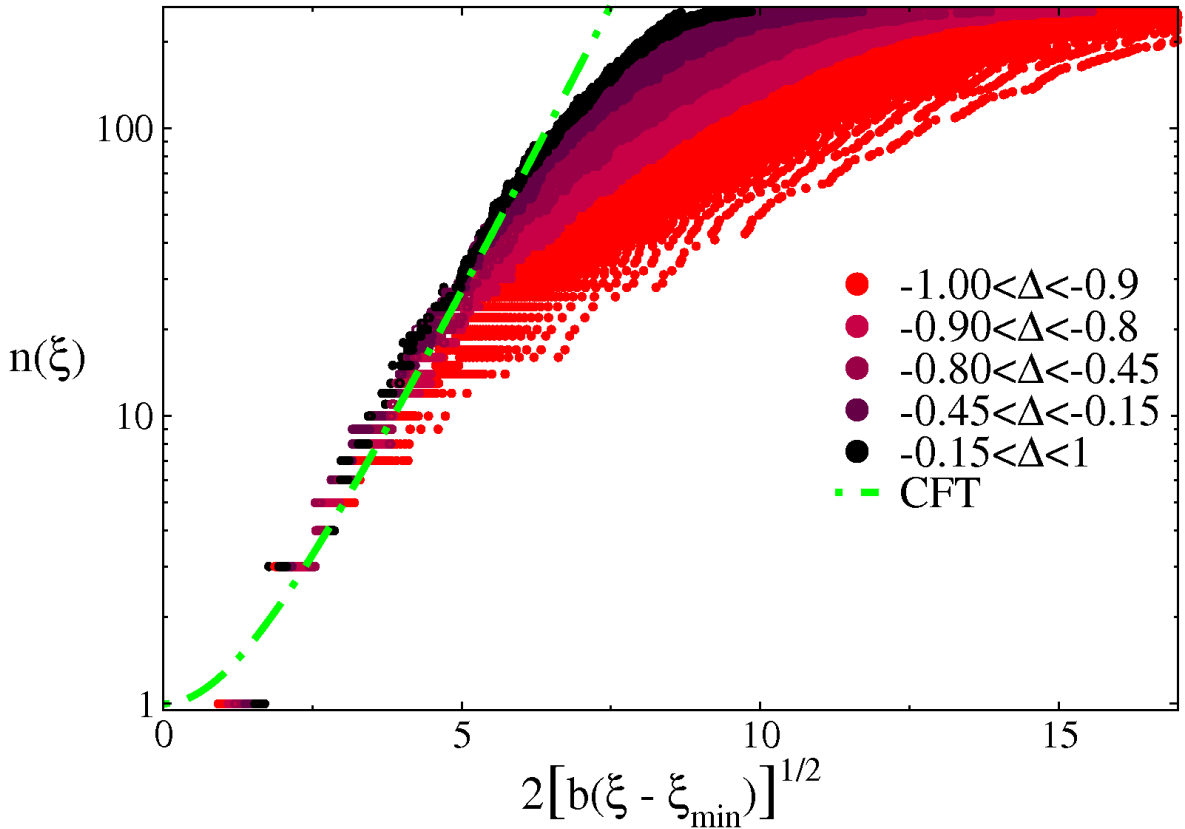}
\end{center}
\caption{ Distribution of the ES for a block with 
 $L_A=8$ embedded in a chain of $L=18$ ($M=9$) sites and
 several values of $\Delta$ (exact diagonalization).
  We denote with $n(\xi)$ the number of 
 ES levels smaller than $\xi$. The  line 
 is the asymptotic CFT result~\eref{CFT}.
 }
\label{CFT_fig}
\end{figure}

In this section we show how the symmetric structure of the $\Delta=-1$ point
affects the distribution of ES levels for $\Delta>-1$.  We use periodic
boundary conditions and consider $M=L/2$ (i.e. the state in the multiplet that
is the ground state at $\Delta>-1$).  It has been shown~\cite{cal-lev-2008}
that for 1D gapless systems described by a conformal field theory (CFT) the
asymptotic distribution of the ES levels is given by
\begin{equation}
\label{CFT}
n(\xi)=I_0(2\sqrt{b(\xi-\xi_{min})})
\end{equation}
where $n(\xi)$ is the number of ES levels smaller than $\xi$, $\xi_{min}$ is
the value of the lowest ES level, $b=c/6\log\big[L\sin(\pi/L L_A)\big]$, $c$
is the central charge ($c=1$ for the $XXZ$ spin chain) and $I_0$ the modified
Bessel function of order zero.  This ES distribution depends only on the
central charge (in this sense is super-universal), i.e.  $\eref{CFT}$ holds
irrespectively of the value of $\Delta$ in the gapless phase.  

However, the ferromagnetic point $\Delta=-1$ (where the model is not conformal
invariant) is the endpoint of the $c=1$ conformal line which describes the
$XXZ$ at $-1<\Delta\le 1$.  Although in the thermodynamic limit (when all the
length scales are sent to infinity) the $CFT$ result is expected to describe
the ES distribution in the whole region $-1<\Delta\le 1$, for a finite chain
deviations from the $CFT$ prediction are expected especially in the vicinity
of $\Delta=-1$.  At the $\Delta=-1$ point itself, the von Neumann and the
Renyi entropies show scale invariance but not conformal
invariance \cite{cas-doy-2011}.  The asymptotic behaviors
are \cite{cas-doy-2011}
\begin{eqnarray}
\label{strange_vnre}
\fl
\nonumber
S_A^{(\gamma)}=\frac{1}{2}\log\Big(\frac{\pi L_A}{2}
\Big)+\frac{\log(\gamma)}{2(\gamma-1)}+{\mathcal O}(1/L_A),
\quad S_A=\frac{1}{2}\log\Big(\frac{\pi L_A}{2}\Big)+
{1\over 2}+{\mathcal O}(1/L_A)\\
\end{eqnarray}
Although the scaling is logarithmic, the prefactors differ from the $CFT$
prediction that would give $S^{(\gamma)}_A=\frac{c}{6}(1+
\frac{1}{\gamma})\log L_A+{\mathcal O}(1)$ and 
$S_A=\frac{c}{3}\log L_A+{\mathcal O}(1)$.

It is interesting to check how this is reflected 
at the level of the ES distribution.
In Fig.~\ref{CFT_fig} we show the ES distribution for 
a chain of $L=18$ and block size $L_A=8$. 
Finite size effects are present, but we still notice a good agreement between
the numerical data and the $CFT$ prediction \eref{CFT} for $\Delta>-0.15$.
The numerical data in the range $-0.15<
\Delta<1$ collapse on the same curve, which is well 
described by the $CFT$ result.  However, in the region $\Delta<-0.15$ no data
collapse is observed and the $CFT$ prediction is a quite poor approximation.
Thus for these block sizes and system sizes the ferromagnetic point
$\Delta=-1$ appears to dominate the physics of the whole region $\Delta<0$.

\section{Conclusions}
\label{conclusions}

In this paper we studied the ES of the spin-$\frac{1}{2}$ $XXZ$ model in the
limit $\Delta\to-1^+$ (ferromagnetic point) for a finite chain with periodic
and open boundary conditions. To this purpose we used exact diagonalization
(with arbitrary precision numerics) and Bethe ansatz. We have characterized
the ES for all the chain eigenstates within the highest spin multiplet
(symmetric multiplet) in the limit $\Delta\to-1^+$.

At $\Delta=-1$ these states are flat superpositions of all the spin
configurations compatible with the state magnetization $S^z_T$.  In other
terms they show a ``mean field''  structure with no notion of distance.
As a consequence the ES at $\Delta=-1$ for each of these eigenstates can be
calculated exactly and shows few levels.  At $\Delta>-1$ the ``mean field''
structure is progressively lost as we move away from the ferromagnetic point.

We summarize our results as follows:

\begin{itemize}
\item [(i)] In the limit $\Delta\to-1^+$, most of the ES levels diverge.   
  For each value of the block magnetization $S^z_A$ there is only one non
  diverging level which reflects components of the state wavefunction that in
  the limit $\Delta\to-1^+$ are flat superposition of all the allowed spin
  configurations.  The divergent behavior is given by
  $\xi=-\alpha\log(\Delta+1)+\beta$ with $\alpha=0,2,4, \dots$ for periodic
  boundary conditions and $\alpha=0, 4,8,\dots$ for open.
\item [(ii)] The multiplicity of an ES level diverging with
  slope $\alpha$ is given by the sequences $m_{per}(\alpha)$ for periodic and
  $m_{open}(\alpha)=m_{per}(\alpha/2)$ for open boundary conditions.   We have
  provided a combinatorial expression [Eq.~\eref{conj}] for these sequences.
\item [(iii)] We have studied the entanglement eigenfunctions, i.e., the
  eigenfunctions of the reduced density matrix $\rho_A$.  The eigenfunctions
  of the non diverging ES levels mimic the behavior of the state wavefunction,
  i.e., in the limit $\Delta\to-1^+$ are flat superpositions of all the
  possible spin configurations.  In contrast, the entanglement eigenfunctions
  of the diverging levels do not become flat but display rich structures. In
  particular, in the sector $M_A=1$ their leading order in $\Delta+1$ can be
  given analytically in terms of the discrete Chebyshev polynomials.  
\item [(iv)] Although in the thermodynamic limit in 
 the whole gapless phase at $-1<\Delta\le 1$ the 
 ES distribution is described by~\eref{CFT}, for
 a finite  chain  significant deviations from the 
 CFT prediction and non universal behavior 
 are observed in the vicinity of $\Delta=-1$. Within the range
 of system sizes considered ($L_A=8$) the ferromagnetic point
 $\Delta=-1$ appears to dominate the whole region
 $\Delta<0$.
\end{itemize}

\newpage
\appendix

\section{Expansion of exact Bethe ansatz solutions around $\Delta=-1$}
\label{bethe_pert}

In this Appendix we outline the Bethe ansatz expansions used in the Article. 

\subsection{Periodic boundary conditions}

In the region $\Delta>-1$ the Bethe equations can be written in the form
\begin{equation}
\label{b_eq}
\arctan\Big(\frac{v_l}{\tan(\zeta/2)}\Big)=-
\frac{\pi}{L}I_l+\frac{1}{L}\sum\limits_{
m\ne l}^{M}\arctan\Big(\frac{v_l-
v_m}{(1-v_l v_m)\tan\zeta}\Big)
\end{equation}
where $M$ is the number of particles in the chain, $v_l$ ($l=1,2,\dots,M$) are
the so called rapidities and $\zeta=\arccos\Delta$. $I_l$ are the Bethe
numbers and each choice of $I_l$ identifies an eigenstate of the $XXZ$
Hamiltonian.  In each $S^z_T$ sector the lowest energy eigenstate (which is
the state in the symmetric multiplet with magnetization $S^z_T$) is identified
by the following choice of the Bethe numbers
\begin{equation}
I_l=-(M+1)/2+l\quad l=1,2,\dots,M
\end{equation}
(Note  $M=L/2-S^z_T$.)  
From the solutions of the Bethe equations the state
can be constructed as
\begin{eqnarray}
\label{bethe_state}
\ket{\psi}=\sum\limits_{{\mathcal P}}\underbrace{
(-1)^{\epsilon_{\mathcal P}}e^{\sum_{l}
k_{{\mathcal P}_l}
x_l+\frac{i}{2}\sum\limits_{l<m}
\Phi(k_{{\mathcal P}_l},k_{{\mathcal P}_m})}
}_{a(x_1,x_2,\dots,x_M)}\ket{x_1,x_2,\dots,x_M}
\end{eqnarray}
where ${\mathcal P}$ runs over all permutations of the set
$\{x_1,x_2,\dots,x_M\}$, $x_l$ being the particle positions.  We denote with
$|x_1,x_2,\dots,x_M\rangle$ the spin configuration with particles (down spins)
at sites $x_1,x_2,\dots$.  The parity of the permutation ${\mathcal P}$ is
denoted by $\epsilon_{\mathcal P}$. We also defined
$k_l\equiv\pi+2\arctan(v_l/
\tan(\zeta/2))
\quad (\textrm{mod}\, 2\pi)$ and the scattering phase
$\Phi(k_l,k_m)\equiv 2\arctan((v_l-v_m)/((1-v_l v_m)
\tan\zeta))$.

The idea of the approach is to solve  the Bethe equations 
\eref{b_eq} perturbatively order by  order 
in $\epsilon\equiv\Delta+1$ imposing a solution 
of the form
$$
v_l=v_l^{(0)}+v_l^{(1)}\epsilon+\dots
$$
to obtain a perturbed expression for the state using \eref{bethe_state}. At
zeroth order the Bethe equations are given by the set of equations
$$
v_l^{(0)}=\frac{2}{L}\sum\limits_{m\ne l}
\frac{1-v_l^{(0)}v_m^{(0)}}
{v_l^{(0)}-v_m^{(0)}}
$$

\subsubsection{First order expansion of the wavefunctions}

Up to first order in $\epsilon$ the amplitude (see \eref{bethe_state}) of the
wavefunction with two particles can be written analytically
\begin{equation}
\label{two_part_bethe}
a(x_1,x_2)=\sqrt{\frac{2}{L(L-1)}}\Big[1+A_2\epsilon-
\frac{\epsilon}{L-1}\big(x_1-x_2+\frac{L}{2}\big)^2\Big]
\end{equation}
where 
$$
A_2=\frac{2}{L(L-1)^2}\sum\limits_{x_1<x_2}\big(x_1-x_2
+L/2\big)^2
$$
ensures the correct first order normalization.  The last term
in \eref{two_part_bethe} shows a quadratic behavior in the inter particle
distance $x_2-x_1$ and linear dependence on $\epsilon$ of the ``curvature'' of
the wavefunction.  Eq.~\eref{two_part_bethe} also shows that the asymptotic
limit $\epsilon\to0$ the wavefunction becomes an equal weight superposition
(no dependence on $x_i$).  These features were shown and discussed in
Figure \ref{two_part_gs}, where we also compared with exact diagonalization
data.

In first order it is possible to go beyond $M=2$ and derive the amplitudes for
any state within the symmetric multiplet:

\begin{equation}
\label{many_part_bethe}
a(x_1,x_2,\dots,x_M)=\frac{1}{\sqrt{{\mathcal I}}}\left[
1+A_M\epsilon ~-~ \frac{\epsilon}{L-1}\sum
\limits_{i<j}^M\left(X_{ij}+\frac{L}{2}\right)^2\right]
\end{equation}
where ${\mathcal I}\equiv 1/M!\prod\limits_{k=0}^{M-1} (L-k)$, $X_{ij}\equiv
x_i-x_j$, and the normalization is
\begin{equation}
A_M=\frac{1}{(L-1){\mathcal I}}\sum\limits_{x_1<x_2<
\cdots<x_M}\sum\limits_{i<j}\big(X_{ij}+\frac{L}{2}\big)^2
\end{equation}
While the $M=2$ wavefunction corresponds to an excited state at $\Delta>-1$,
the formula above includes as a particular case ($M=L/2$) the wavefunction of
the ground state in the gapless phase. One important property
of \eref{many_part_bethe} is that it is quadratic and symmetric in all the
interparticle distances $x_i-x_j$.  An example of wavefunction amplitudes
obtained from~\eref{many_part_bethe} is shown in Fig.~\ref{two_part_gs} ({\bf
c}).

\subsection{Open boundary conditions}

It is possible to treat the open boundary conditions
case within the Bethe ansatz formalism~\cite{alc-87}.
The Bethe equations now read

\begin{equation}
\fl
k(v_l)=\frac{\pi}{L+1}I_l-\frac{1}{L+1}\Phi(v_l,-v_l)  ~-~  
\frac{1}{2}\frac{1}{L+1}\sum\limits_{m\ne l}[\Phi(v_l,-v_m)+\Phi(v_l,v_m)]
\end{equation}

with $\Phi(v_l,v_m)\equiv 2\arctan
((v_l-v_m)/((1-v_l v_m)
\tan\zeta))$ and $k(v_l)\equiv\pi+2
\arctan(v_l\tan(\zeta/2))$. 
The Bethe numbers giving the lowest energy state 
in each magnetization sector (i.e. the states 
within the symmetric multiplet) are now $I_l=
\{1,2,\dots,M\}$. The main difference with respect to 
the periodic boundary conditions  is the presence
of the ``reflected'' phase shifts $\Phi(v_l,-
v_m)$.

This fact has dramatic consequences in the 
wavefunction and therefore in the ES. The general 
expression of the chain wavefunction has the form

\begin{equation}
\fl
\label{open_bethe_state}
\ket{\psi}=\sum\limits_{{\mathcal P}}
(-1)^{\epsilon_{{\mathcal P}}}A(v_{{\mathcal P}_1},
v_{{\mathcal P}_2},\dots,v_{{\mathcal P}_M})
e^{i(k_{{\mathcal P}_1}x_1+k_{{\mathcal P}_2}
x_2+\cdots +k_{{\mathcal P}_M}x_M)}
|x_1,x_2,\cdots,x_M\rangle
\end{equation}

where the sum is over all permutations and 
(arbitrary number of) reflections 
of the  set of rapidities $\{v_1,v_2,\cdots,v_M\}$.
The amplitude $A(v_{{\mathcal P}_1},
v_{{\mathcal P}_2},\dots,v_{{\mathcal P}_M})
$ is not as simple as for periodic boundary
and we do not show it.
As for periodic boundary conditions we can
expand the Bethe equations in the vicinity of 
$\Delta=-1$. Quite interestingly now we find that
 the lowest order solutions are determined by the 
equations

\begin{equation}
\frac{1}{v_l^{(0)}}=\frac{1}{L+1}\frac{1+
(v_l^{(0)})^2}{v_l^{(0)}}+\\
\frac{1}{L+1}\sum\limits_{m\ne l}\Big[
\frac{1-v_l^{(0)}v_m^{(0)}}{v_l^{(0)}-
v_m^{(0)}}+\frac{1+v_l^{(0)}v_m^{(0)}}
{v_l^{(0)}+v_m^{(0)}}\Big]
\end{equation}

By expanding~\eref{open_bethe_state} in $\Delta+1$ we find quite remarkably
that the sum over the reflections of the rapidities $v_l\to-v_l$ cancel at
first order in $\Delta+1$ the products of the form $x_ix_j$.  
This fact leads to the suppression of the first order ES levels for 
open boundary conditions, as we have seen (Figures \ref{spec_intro} 
and \ref{open_many_part_comb}).We provide as an example of this cancellation 
the first order amplitude $a(x_1,x_2)$ for the (unnormalized) chain 
wavefunction in the case with two particles, which reads
\begin{equation}
\fl
a(x_1,x_2)=1-\frac{\epsilon}{L}\left[L^2/2+3-L(1/6+x_1+3x_2)
+2x_1(x_1+1)+2x_2(x_2+1)\right]
\end{equation}
Our numerical results indicate that the same type of cancellation will happen at
every odd order, but for higher orders this is too cumbersome to perform
explicitly.

\end{document}